\documentclass[pdflatex,iicol,sn-mathphys-num]{sn-jnl}
\usepackage{amsmath,amssymb,amsfonts}
\newcommand{\ignore}[1]{}
\newcommand{\THGEE}{\operatorname{TH\mbox{-}GEE}}
\usepackage{booktabs}
\usepackage{graphicx}
\usepackage{multirow}
\usepackage{array,color,xcolor}
\usepackage{xurl}
\usepackage[nopatch=footnote]{microtype}
\usepackage{setspace}

\begin{document}

\title{Measuring Time-Horizon Engagement Effectiveness: Persistence, Recency, and Re-Emergence}

\author*[1,2]{\fnm{Bonnie} \sur{Rushing}}
\email{brushing@uccs.edu}

\author[1,2]{\fnm{Ethan} \sur{Anderson}}
\email{eander34@uccs.edu}

\author[2]{\fnm{William} \sur{Hersch}}
\email{william.hersch@afacademy.af.edu}

\author[1]{\fnm{Shouhuai} \sur{Xu}}
\email{sxu@uccs.edu}

\affil[1]{\orgdiv{Laboratory for Cybersecurity Dynamics, Department of Computer Science},
  \orgname{University of Colorado Colorado Springs},
  \orgaddress{\city{Colorado Springs}, \state{CO}, \country{USA}}}

\affil[2]{\orgname{United States Air Force Academy},
  \orgaddress{\city{Colorado Springs}, \state{CO}, \country{USA}}}

\affil[]{\textbf{Google Scholar profile:}
\url{https://scholar.google.com/citations?user=aL2gUUMAAAAJ&hl}}

\abstract{Online attention is commonly summarized using cumulative volume, peak activity, or arithmetic averages, but such measures can obscure differences between activity that is sustained over time, concentrated near the present, or renewed after dormancy. This paper introduces the Time-Horizon Engagement Effectiveness (TH-EE) framework, which constructs interpretable temporal profiles of online attention by distinguishing three related but non-equivalent properties: persistence, recency, and re-emergence. We evaluate the framework through controlled engagement traces, a proof-of-concept application to 1,850 YouTube videos across 37 topics (18 cohorts selected a priori as exemplars of persistent, acute, cyclical, and recently originating attention, and 19 cohorts corresponding to authoritatively debunked claims), and an event-level validation on eleven years of daily Wikipedia pageview series for the same topics. The controlled analyses show that the framework distinguishes distributed activity from concentrated bursts, introduces temporal-order sensitivity through recency weighting, and identifies renewed activity after a defined dormant interval. On the event-level series, the framework's reactivations co-locate with Kleinberg burst onsets, and PELT change points far more often than chance. The YouTube application shows that debunked-claim cohorts do not occupy a unique region of temporal-profile space; they exhibit heterogeneous patterns that overlap substantially with benign topics. These results support treating persistence, recency, and re-emergence as separate dimensions of online attention. TH-EE is a descriptive and comparative measurement framework, not a classifier of misinformation, coordination, intent, or content veracity.}


\maketitle

\textit{The views expressed are those of the authors and do not reflect the official policy or position of the U.S. Air Force, Department of Defense, or U.S. Government.}

\section{Introduction}
\label{sec:introduction}

Online attention is not distributed uniformly through time. Some topics receive a large but short-lived burst of engagement, whereas others remain visible at moderate levels across extended periods. Still, others appear to become dormant before returning in response to a new event, renewed publication, adaptation, recommendation, or public discussion. These trajectories may have similar cumulative engagement totals while representing substantially different temporal conditions.

Research on collective attention has documented novelty decay, endogenous and exogenous activity, heavy-tailed diffusion, emotional transmission, and heterogeneous cascade structures \cite{wu2007novelty,crane2008robust,lehmann2012dynamical,
bakshy2011everyone,brady2017emotion}.
Nevertheless, many empirical studies continue to summarize engagement using totals, peaks, arithmetic averages, or other volume-dominated measures. Such summaries answer how much engagement occurred, but they do not necessarily show whether activity was broadly distributed, concentrated near the end of an observation period, or renewed following dormancy.

These distinctions are relevant across many computational social science applications. Public health communication, political discussion, entertainment, scientific controversies, social movements, conspiracy theories, and misinformation may all exhibit persistent, acute, cyclical, or re-emergent attention. Repeated exposure can affect familiarity and perceived accuracy \cite{pennycook2018prior}, but persistence itself is not evidence of falsity, persuasion, coordination, or malicious intent. Educational resources, institutional communication, seasonal events, and durable communities may produce similar temporal patterns.

This paper treats temporal structure as an object of measurement rather than as a proxy for content type. We introduce the Time-Horizon Engagement Effectiveness (\textit{TH-EE}) framework, which represents the temporal profile of a topic, narrative, campaign, hashtag, or related collection of content. We develop a geometric implementation, termed Time-Horizon Geometric Engagement Effectiveness (\textit{TH-GEE}), for settings in which engagement values are nonnegative, highly skewed, and potentially dominated by a small number of extreme observations.

The framework distinguishes three dimensions:

\begin{enumerate}
    \item \emph{Persistence}: the extent to which observable activity is     distributed across an observation horizon rather than confined to a small    number of windows;

    \item \emph{Recency}: the extent to which engagement is concentrated near    the end of the observation horizon; and

    \item \emph{Re-emergence}: the return of activity after prior activity and a qualifying interval of dormancy.
\end{enumerate}

These dimensions are related but not interchangeable. Persistent activity need not be recent, recent activity may represent a first appearance rather than a return, and continuously active content may have high persistence and recency without ever becoming dormant.

The paper makes three contributions.
First, it contributes a new methodological result for computational social science: a measurement framework that decomposes the temporal structure of online attention into three separable, interpretable dimensions---persistence, recency, and re-emergence---applicable to any topic, narrative, or campaign observable as a windowed engagement trace, independent of platform.
Second, it provides transparent measures for operationalizing these dimensions over discrete time windows. 
Third, it evaluates the framework through controlled construct tests, a proof-of-concept analysis of 37 YouTube topics (1,850 videos), and an event-level validation on eleven years of daily Wikipedia pageview series for the same topics, comparing the framework's reactivations against burst detection and change-point estimation.

The results support a deliberately limited conclusion. Temporal measures can describe whether observable attention is persistent, recent, dormant, or re-emergent, but they cannot independently establish \textit{why} that pattern occurred. Content veracity, persuasion, intent, coordination, and causal mechanisms require additional evidence.

\subsection{Scope and Interpretation}
\label{sec:scope}

TH-EE characterizes the temporal structure of observable online activity. 
We retain the term ``engagement effectiveness" from \cite{rushing2026engagement}, where it denotes engagement relative to transmissions.
It does not infer the beliefs, intentions, or psychological states of the people who produce or interact with content. Engagement (e.g., likes, comments, shares, publication) may represent endorsement, disagreement, correction, curiosity, entertainment, or coordinated amplification. Accordingly, a high score on any TH-EE component should not be interpreted as evidence that content is false, persuasive, malicious, coordinated, or artificially generated.
TH-EE is a measurement framework rather than a causal model or content classifier. Its outputs may be used as descriptive variables in subsequent statistical, comparative, or predictive analyses, but substantive interpretation requires evidence appropriate to the research question.

\section{Related Work}
\label{sec:related}
We divide related prior studies into three categories.

\subsection{Temporal dynamics of online attention}\label{sec:attention}

Collective attention commonly rises and decays rather than remaining stationary. Wu and Huberman \cite{wu2007novelty} model the rapid decay of attention to novel online content, while Crane and Sornette \cite{crane2008robust} distinguish endogenous and exogenous (internal and external) response patterns using temporal activity profiles. 
Research on Twitter further identifies multiple dynamical classes, including sudden exogenous events and slower endogenous growth \cite{lehmann2012dynamical}. These findings caution against treating cumulative engagement as a sufficient description of social response.
These studies motivate TH-EE by establishing that attention trajectories differ in their timing and generative processes. TH-EE does not attempt to infer the causal process that generated a trace; instead, it provides a compact descriptive profile of whether engagement is distributed across the horizon, concentrated near its end, or renewed after dormancy. It therefore translates insights from temporal-dynamics research into interpretable measures suitable for comparing many topic-level patterns.

Burst detection and point-process research provide related approaches to temporal structure. Kleinberg's burst model identifies intervals in which event rates depart from a baseline state \cite{kleinberg2003bursty}; Hawkes processes model self-exciting event sequences in which prior events increase the short-term likelihood of subsequent events \cite{hawkes1971spectra}; and change-point methods estimate transitions between statistical regimes \cite{killick2012optimal}. These methods are more expressive than a scalar engagement summary, but they often require event-level sequences, model fitting, or assumptions that may be difficult to satisfy in large-scale triage settings. TH-GEE is positioned as an interpretable summary that can complement, rather than replace, these sequence-oriented approaches.

\subsection{Influence, repetition, and engagement}\label{sec:influence}

Online influence is broader than misinformation and does not necessarily imply manipulation. Nevertheless, strategically produced or repeated content can shape exposure, salience, and perceived credibility \cite{lazer2018science,wardle2017information,pennycook2018prior}. AI-assisted generation further reduces the cost of producing message variants and maintaining repeated communication, increasing the practical importance of longitudinal measurement \cite{drolsbach2025characterizing,rushing2026defining}. Artifact-based methods such as watermarking, provenance, media forensics, and text classification address whether content may be synthetic \cite{kirchenbauer2023watermark,c2pa2023spec,verdoliva2020media,mitchell2023detectgpt}; they do not describe how engagement with that content is distributed through time.
This literature defines the broader detection context in which TH-EE may be used in the future. Content/artifact-based methods characterize what an item is or how it was produced; TH-EE characterizes how engagement evolves over time. 
The approaches are complementary: temporal profiles can enrich a detector or analyst workflow but cannot substitute for evidence about content, provenance, coordination, or intent.

Engagement effectiveness provides a behavioral measure based on weighted interactions relative to transmissions \cite{rushing2026engagement}. Extending this measure across windows creates a trace suitable for temporal comparison. Because online interactions are commonly heavy-tailed, arithmetic summaries can be dominated by exceptional observations \cite{bakshy2011everyone,limpert2001lognormal}. A geometric summary limits the effect of extreme values and lowers the score when many time windows have little activity. This helps measure persistence, but not the order or timing of engagement.
TH-EE extends this window-level engagement effectiveness measure into a longitudinal framework: Equation~\ref{eq:window_engagement} supplies each window's value, and the three components summarize complementary properties of the resulting sequence.

\subsection{Temporal signatures in misinformation and cognitive attack detection}
\label{sec:temporal_classification}

Temporal information has frequently been used to distinguish misinformation, coordinated influence, and other anomalous online activity from ordinary diffusion.
Research on false news diffusion supports the proposition that temporal behavior can be informative. False and true information may differ in diffusion speed, reach, cascade depth, and the roles played by human and automated accounts. Other work uses burst detection, change-point analysis, and self-exciting point processes to identify statistically unusual periods of activity \cite{kleinberg2003bursty,hawkes1971spectra,killick2012optimal}. These methods demonstrate that temporal features can contribute to detection and characterization. 

A relevant expectation is that influence narratives may follow a recursive or re-emergent trajectory. An actor introduces a narrative, allows attention to decay or become dormant, and later reintroduces the claim through new message variants, triggering events, reposting, or amplifier accounts. Re-emergence is plausible because influence activity may be adapted to changing events and repeated exposure can increase familiarity and perceived accuracy \cite{pennycook2018prior}. It is therefore reasonable to treat renewed activity after dormancy as an important pattern for analyst triage.

The proposition that re-emergence is the temporal signature of cognitive attacks or misinformation is less well supported. Prior studies have identified differences in the speed, depth, persistence, and recurrence of false and misleading information, but they do not establish a single trajectory shared by all misinformation \cite{delvicario2016spreading,vosoughi2018spread,shin2018diffusion,kauk2025reappearance}. 
Observed diffusion reflects the interaction of content, audience interest, network structure, platform visibility, external events, countermessaging, and data-collection choices \cite{hodas2014simple,shin2018diffusion}. An adversarial campaign may consequently appear as an acute burst, continuous low-intensity activity, periodic recurrence, or re-emergence. 
Furthermore, recurrence is not unique to false or adversarial content. Ordinary online attention can also be shaped by exogenous events, endogenous growth, novelty decay, periodic interest, and renewed visibility \cite{wu2007novelty,crane2008robust,lehmann2012dynamical}. Re-emergence is therefore neither necessary nor sufficient evidence of misinformation or coordinated influence.

\smallskip

The preceding thrusts of research on temporal signatures inspired the design of TH-GEE as follows. Rather than treating persistence, recency, or re-emergence as class labels for misinformation, TH-GEE measures them as separable patterns that may occur in both benign and misleading content.
This guides us to design our empirical study questions to investigate whether temporal measures can distinguish the multiple trajectories that misinformation may follow, rather than whether all misinformation follows a recursive trajectory.

\section{The TH-EE Framework}
\label{sec:framework}

The TH-EE framework uses different methods to combine engagement across time windows. The framework takes as input a given {\em unit of analysis}, denoted by \(c\), such as a campaign, narrative, hashtag, topic, or group of related posts. The framework operates in a discrete-time model by dividing the given observation time horizon into \(T\) consecutive time windows, and a time window may be an hour, day, week, or month, depending on the application. 
Table~\ref{tab:variables} summarizes the notations used throughout the paper for the framework. The framework consists of four steps.

\begin{table*}[!htbp]
\centering
\caption{TH-EE framework notations.}
\label{tab:variables}
\small
\renewcommand{\arraystretch}{1.16}
\begin{tabular}{p{2.5cm}p{12.2cm}}
\toprule
\textbf{Notation} & \textbf{Meaning} \\
\midrule
$c$ & Unit of analysis (e.g., a campaign, narrative, hashtag, topic, related set of posts) \\
$h$ & Observation horizon \\
$T$ & Number of consecutive time windows in horizon $h$ \\
$\tau,r,s$ & Time-window indices in $\{1,\ldots,T\}$ \\
$j$ & Interaction-type index \\
$m$ & Number of interaction types \\
$i_{j,\tau}(c)$ & Number of type-$j$ interactions associated with $c$ in window $\tau$ \\
$w_j$ & Weight assigned to interaction type $j$ \\
$t_{\tau}(c)$ & Number of transmissions or posts associated with $c$ in window $\tau$ \\
$E_{\tau}(c)$ & Engagement score for $c$ in window $\tau$ \\
$\mathbf{E}_h(c)$ & Ordered sequence of window-level engagement scores over horizon $h$ \\
$P_h(c)$ & Geometric Persistence Score over horizon $h$ \\
$\pi$ & Permutation of the window indices \\
$\gamma$ & Recency-decay parameter \\
$\alpha_{\tau}(\gamma)$ & Normalized recency weight assigned to window $\tau$ \\
$\THGEE_h(c;\gamma)$ & Recency-weighted TH-GEE score over horizon $h$ \\
$q_h$ & Activity threshold used to classify a window as active \\
$d$ & Minimum number of consecutive inactive windows defining dormancy \\
$z_{\tau}(c;q_h)$ & Binary active-window indicator \\
$r_{\tau}(c;q_h,d)$ & Binary indicator that window $\tau$ is a qualifying reactivation \\
$R_h(c;q_h,d)$ & Re-Emergence Index over horizon $h$ \\
$\epsilon$ & Small positive constant preventing division by zero \\
$\boldsymbol{\Phi}_h(c)$ & Three-component temporal engagement profile \\
\bottomrule
\end{tabular}
\end{table*}

\subsection{Step 1: Calculate engagement score within each time window}

For each time window $\tau$, we first calculate a window-level engagement score by combining observable interaction types, such as views, likes, comments, reposts, or shares, and normalizing the weighted total by the number of posts or transmissions produced during that window.

Let $i_{j,\tau}(c)$ be the number of interactions of type $j$ that are attributable to unit $c$ during window $\tau\in\{1,\ldots,T\}$; for example, these may be interactions with posts belonging to a specified influence campaign or topic cohort. 
Let $m$ be the total number of interaction types, 
$w_j$ the weight assigned to type $j$, and 
$t_{\tau}(c)$ the number of posts or transmissions associated with $c$ during window $\tau$.
Then, the engagement score is defined as:
\begin{equation}
E_{\tau}(c)=
\begin{cases}
\displaystyle
\frac{\sum_{j=1}^{m} w_j i_{j,\tau}(c)}
     {t_{\tau}(c)},
& t_{\tau}(c)>0,\\[8pt]
0, & t_{\tau}(c)=0.
\end{cases}
\label{eq:window_engagement}
\end{equation}

The engagement score for time window $\tau$ is zero if no campaign content is transmitted during the window. Then, we obtain a sequence of engagement scores:
\begin{equation}
\mathbf{E}_{h}(c)
=
\left(
E_1(c),E_2(c),\ldots,E_T(c)
\right).
\label{eq:trace}
\end{equation}

A sequence $(20,18,16,15,14)$ shows engagement continuing across all five time windows, while a sequence $(83,0,0,0,0)$ shows one large burst followed by no activity.
The choice of time window length affects the results. Short windows preserve brief bursts but often contain more zeros, and longer windows are smoother but may hide short dormant periods or rapid reactivation \cite{wu2007novelty,lehmann2012dynamical}. Thus, results should be tested using multiple window lengths to ensure robustness.

\subsection{Step 2: Calculate persistence score within time horizon}

We propose measuring whether engagement is spread across the observation period rather than concentrated in only a few time windows via the following \emph{Geometric Persistence Score}:
\begin{equation}
P_h(c)=
\exp\left[
\frac{1}{T}
\sum_{\tau=1}^{T}
\log\left(1+E_{\tau}(c)\right)
\right]-1,
\label{eq:persistence}
\end{equation}
which is a geometric mean written in logarithmic form. The $+1$ shift permits zero-valued windows, which would otherwise force the entire geometric mean to zero \cite{delacruz2018geometric}; the logarithmic form remains numerically stable for long traces and large scores. The geometric mean itself is not new---our contribution is using it to measure persistence alongside recency and re-emergence. By design, $P_h$ combines temporal coverage with engagement magnitude: a trace that is both broadly active and highly engaged scores higher than one that is merely broadly active. 
When distribution must be isolated from scale, the analysis pairs $P_h$ with the scale-free active-window coverage indicator, as in Sections \ref{sec:res_observational} and \ref{sec:res_debunked}.

The definition has two salient features. First, it does not consider the order of the windows. For example, two sequences of engagement scores 
$(20,10,5,0)$
and $(0,5,10,20)$
lead to the same persistence score. This highlights the desired pattern that \(P_h\) measures how engagement is distributed across windows, but not whether engagement occurred early or recently.
This pattern can be written formally as
\begin{equation}
\begin{split}
P_h(c) &= P_h(E_1(c),\ldots,E_T(c)) \\
       &= P_h\bigl(E_{\pi(1)}(c),\ldots,E_{\pi(T)}(c)\bigr),
\end{split}
\label{eq:permutation}
\end{equation}
where \(\pi\) is a permutation defined on $(1,\ldots,T)$. 

Second, the geometric aggregation function is advantageous over other aggregations, such as the algebraic mean. In particular, it limits the influence of isolated extreme engagement scores and lowers the score when many time windows contain small engagement scores
\cite{limpert2001lognormal}.

\subsection{Step 3: Calculate recency weight and recency-weighted engagement effectiveness score within time horizon}
\label{sec:thgee}

Since persistence alone cannot determine whether a unit of analysis (e.g., campaign) is still active or not, we propose measuring recency via parameter $\gamma\geq0$, which gives a higher importance to interactions in the latest time windows (i.e., the most recent time window with respect to the time horizon or observation time window $T$ receives the largest weight) and lower importance to older time windows. We define the {\em recency weight} $\alpha_\tau(\gamma)$ associated with each time window $\tau$ as follows:
\begin{equation}
\alpha_{\tau}(\gamma)=
\frac{\exp[-\gamma(T-\tau)]}
{\sum_{r=1}^{T}\exp[-\gamma(T-r)]}.
\label{eq:weights}
\end{equation}
The weights are positive and normalized to sum to 1. Parameter \(\gamma\) controls how quickly older interactions lose importance: when \(\gamma=0\), all time windows receive the same recency weight; a small positive \(\gamma\) gives recent time windows slightly higher recency weight; a larger \(\gamma\) places much greater emphasis on the most recent time windows.

Then, we define the \emph{Time-Horizon Geometric Engagement Effectiveness score} within time horizon, denoted by \(\THGEE_h(c;\gamma)\), as follows:
\begin{equation}
\begin{split}
\operatorname{TH\mbox{-}GEE}_h(c;\gamma)= \\
\exp\left[
\sum_{\tau=1}^{T}
\alpha_{\tau}(\gamma)
\log\left(1+E_{\tau}(c)\right)
\right]-1.
\end{split}
\label{eq:thgee}
\end{equation}

Like $P_h$, $\THGEE_h(c;\gamma)$ is unbounded and expressed on the original engagement scale rather than as a normalized index; this preserves magnitude information within a trace, while cross-topic comparisons rely on scale-free quantities such as the ratio $\THGEE_h/P_h$ (Section~\ref{sec:evaluation}).

Unlike the unweighted persistence score $P_h(c)$, \(\THGEE_h(c;\gamma)\) changes when engagement moves from an early time window or small $\tau$ to a recent time window or large $\tau$. 

For example, the engagement trace $\mathbf{E}_{h}(c)=(20,10,5,0)$ will generally receive a lower TH-GEE score than the engagement trace $\mathbf{E}_{h}(c)=(0,5,10,20)$ because the same engagement values occur in \textit{more recent windows} in the second trace.
When \(\gamma=0\), every window receives equal weight, the TH-GEE score degenerates to the Geometric Persistence Score, namely:
\begin{equation}
\operatorname{TH\mbox{-}GEE}_h(c;0)=P_h(c).
\label{eq:reduction}
\end{equation}
Thus, the relationship between these two metrics can be understood as follows.
On one hand, \(P_h\) asks {\em was engagement sustained across the observation time horizon?} On the other hand, $\operatorname{TH\mbox{-}GEE}_h$ asks \emph{was engagement sustained and still active near the end of the observation time horizon?}

\subsection{Step 4: Calculate Re-Emergence Index within time horizon}
\label{sec:reemergence}

We observe that recency and re-emergence are not the same thing. For instance, a campaign may remain active throughout the observation time horizon and therefore have a high \(\THGEE_h(c;\gamma)\) without ever disappearing. Thus, we propose defining the \emph{Re-Emergence Index}, denoted by \(R_h(c;q_h,d)\), as the share of the trace's log-scaled engagement that occurs at qualifying reactivation points.

To define \(R_h(c;q_h,d)\), we observe that
re-emergence requires three components, using an influence campaign as an example of a unit of analysis:
(i) the campaign is active; (ii) the campaign becomes inactive for a period; and (iii) the campaign becomes active again. To identify these components, we first need to classify each time window as active or inactive with respect to an engagement score threshold \(q_h\):
\begin{equation}
z_{\tau}(c;q_h)
=
\mathbb{I}\left[E_{\tau}(c)\geq q_h\right],
\label{eq:active}
\end{equation}
where $\mathbb{I}[\cdot]$ is the indicator function, $z_{\tau}(c;q_h)=1$ means time window $\tau$ is active and $z_{\tau}(c;q_h)=0$ otherwise.

\ignore{
The indicator
\(\mathbb{I}[\cdot]\) equals one when the condition is true and zero
when it is false. Therefore,

\[
z_{\tau}=
\begin{cases}
1, & \text{if the window is active},\\
0, & \text{if the window is inactive}.
\end{cases}
\]
}

For example, suppose \(q_h=5\). A sequence of engagement score
$(12,9,2,1,0,11)$
becomes the activity sequence 
$(1,1,0,0,0,1)$.
This sequence shows initial activity, three inactive time windows, and then
renewed activity.
Let \(d\) be the minimum number of consecutive inactive windows needed
to declare dormancy. If \(d=3\), the example above contains a qualifying
dormant period.
A reactivation occurs when (i) the current window is active; (ii) the previous \(d\) windows were inactive; and (iii) the campaign was active at least once before the dormant period. 
In general, the reactivation indicator can be written as:
\begin{equation}
\begin{split}
    r_{\tau}(c;q_h,d)= \\
z_{\tau}
\left(\prod_{j=1}^{d}(1-z_{\tau-j})\right)
\mathbb{I}
\left[
\sum_{s=1}^{\tau-d-1}z_s>0
\right],
\end{split}
\label{eq:reactivation}
\end{equation}
which can be understood via its three parts explained below:
\begin{itemize}
\item \(z_{\tau}\), which indicates (e.g.) that the campaign is active or not in the current time window $\tau$;

\item \(\prod_{j=1}^{d}(1-z_{\tau-j})\), which indicates each of the previous \(d\) windows was inactive;

\item \(\mathbb{I}
[\sum_{s=1}^{\tau-d-1}z_s>0]\): which indicates the campaign was active before the dormant period. Note that this condition is important because it prevents the first observed appearance of a campaign from being incorrectly labeled as re-emergence.
\end{itemize}

Next, we define the Re-Emergence Index \(R_h(c;q_h,d)\), which aims to measure how much of the campaign's engagement occurs at the reactivation points, as follows:
\begin{equation}
\begin{split}
    R_h(c;q_h,d)= \\
\frac{
\sum_{\tau=d+2}^{T}
r_{\tau}(c;q_h,d)
\log\left(1+E_{\tau}(c)\right)
}{
\sum_{\tau=1}^{T}
\log\left(1+E_{\tau}(c)\right)+\epsilon
},
\end{split}
\label{eq:reemergence}
\end{equation}
where the numerator measures engagement occurring during identified reactivations, the denominator measures engagement across the complete trace, and the small  \(\epsilon>0\) is used to prevent division by zero when no engagement is observed. $R_h$ deliberately credits only the reactivation window itself rather than the complete renewed episode: activity that continues beyond the first returned window is ordinary continuation and is already reflected in $P_h$ and $\THGEE_h$.

We observe that a high \(R_h\) value means that a substantial portion of the campaign's engagement occurred when activity returned after dormancy, and that a low value means engagement was either continuous, never returned, or was mostly concentrated outside the reactivation windows.

\subsection{Temporal engagement profile}
\label{sec:profile}
We do not advocate combining the three metrics into a single scalar because doing so would obscure their distinct meanings and introduce additional weights requiring empirical justification.
Instead, we propose representing them as a {\em temporal engagement profile}, defined as:
\begin{equation}
\begin{split}
\boldsymbol{\Phi}_h(c)= 
\left(
P_h(c),
\operatorname{TH\mbox{-}GEE}_h(c;\gamma),
R_h(c;q_h,d)
\right).
\end{split}
\label{eq:profile}
\end{equation}

\begin{table}[!htbp]
\caption{Temporal constructs in the TH-EE framework.}
\label{tab:constructs}
\centering
\small
\begin{tabular}{p{0.18\columnwidth}>{\raggedright\arraybackslash}p{0.4\columnwidth}>{\raggedright\arraybackslash}p{0.25\columnwidth}}
\toprule
\textbf{Construct} & \textbf{Analytical question} & \textbf{Measure} \\
\midrule
Persistence &
Is engagement distributed across multiple windows rather than concentrated in a brief burst? &
\textit{Geometric Persistence Score} $P_h(c)$ \\
Recency &
Is engagement concentrated near the time at which the trace is assessed? &
\textit{TH-GEE score} $\operatorname{TH\mbox{-}GEE}_h(c;\gamma)$ \\
Re-emergence &
Does activity return after prior activity and at least $d$ consecutive dormant windows? &
\textit{Re-Emergence Index} $R_h(c;q_h,d)$ \\
\bottomrule
\end{tabular}
\end{table}

Table~\ref{tab:constructs} further summarizes the analytical role of each component in $\boldsymbol{\Phi}_h(c)$.
The profile can describe, for instance, a briefly viral narrative, a continuously active narrative, an inactive narrative, and a narrative that has recently returned. 
For instance, an acute sequence of engagement may have high volume but low persistence; a continuously active sequence of engagement may have high persistence and high recency but low re-emergence; a once-active but now-dormant sequence of engagement may retain a moderate persistence score while receiving a low TH-GEE score; a sequence of engagement that returns after dormancy may have elevated TH-GEE and $R_h$, depending on the magnitude and timing of renewed engagement. 

\section{Empirical Study}
\label{sec:design}

In this section, we present an empirical study to demonstrate the usefulness of the framework.

\subsection{Dataset and Sampling}
\label{sec:dataset}

We use public metadata collected through the YouTube Data API v3. The dataset contains 1,850 videos returned by searches for 37 topic terms. For each search term, we retain up to 50 returned videos and record each videos' publication timestamp, like count, and comment count at the time of collection.
The dataset contains two distinct sets. The first consists of 18 topics selected \emph{a priori} as illustrative benign examples of persistent, acute, cyclical, or recently originating attention. The second consists of 19 search terms associated with claims that have been authoritatively debunked (e.g., misinformation and conspiracy theories). These cohorts are used to examine whether attention to debunked claims exhibits a common (benign) temporal profile.

The resulting cohorts are search-based samples rather than probability samples. They do not constitute a census of all videos related to a topic, and the returned videos may be affected by query wording, search relevance, platform indexing, content availability, and the time at which data were collected. Accordingly, results describe the retrieved cohorts and should \textit{not} be interpreted as estimates of all YouTube content or audience engagement concerning those topics.

In addition to the YouTube corpus, we retrieve daily pageview series from the Wikimedia REST API for the English Wikipedia article corresponding to each topic; all 37 topics have one. These series record audience interactions dated to the day they occurred, covering 2015-07-01 to 2026-06-30, and serve as the event-level data source for the validation in Section~\ref{sec:res_eventlevel}.

\subsection{Unit of Analysis and Trace Construction}
\label{sec:sequence}

The unit of analysis is the topic-search cohort, defined as the set of videos returned for a specified search term. Each video is assigned to the 90-day (quarterly) window containing its publication timestamp. Within each window, the interaction values associated with videos published during that period are aggregated to produce a publication-cohort engagement value.

For the YouTube application, likes and comments are assigned weights of \(w_{\mathrm{like}}=0.3\) and \(w_{\mathrm{comment}}=0.7\). Share counts are not available through the API and are therefore excluded. The window-level score is calculated using Equation~\ref{eq:window_engagement}. Each video additionally contributes a unit publication weight ($w_{\mathrm{pub}} = 1$) which distinguishes windows containing publications that received no interactions from windows with no publications.

The resulting trace must be interpreted carefully. YouTube provides video publication timestamps and cumulative interaction counts observed at collection time, but not timestamped histories of when likes and comments occurred. A quarterly value therefore represents the eventual observed engagement associated with videos published during that quarter. It does not represent all audience interactions that occurred during the quarter.

Consequently, persistence in this application refers to the distribution of engaged publication cohorts across time. Recency refers to whether videos with high eventual engagement were published near the end of the observation horizon. Reactivation refers to renewed publication-cohort activity following a qualifying interval without active publication cohorts. Direct measurement of audience-level persistence or re-engagement would require repeated engagement snapshots or event-level interaction records.

\subsection{Research questions}\label{sec:rqs}

The empirical study addresses seven questions:
\begin{itemize}
\item RQ1: Can $P_h$ distinguish engagement distributed across many windows from matched-volume engagement concentrated in an acute burst?
\item RQ2: Does recency-weighted TH-GEE change when the same multiset of window-level engagement values is assigned to different temporal positions?
\item RQ3: Can $R_h$ distinguish continuous activity from renewed activity following a specified dormant interval?
\item RQ4: How sensitive are the three measures to window length, decay coefficient $\gamma$, activity threshold $q_h$, and dormancy duration $d$?
\item RQ5: Do the temporal profiles of real YouTube topics reflect their expected persistent, acute, cyclical, and recent patterns?
\item RQ6: Do debunked-claim (e.g., misinformation) topics share a common temporal signature, or do they exhibit heterogeneous temporal profiles?
\item RQ7: Do the reactivations identified by $R_h$ on event-level series co-locate with events found by sequence-oriented detectors, and do publication-cohort profiles agree with event-level profiles for the same topics?
\end{itemize}

\subsection{Evaluation Design}
\label{sec:evaluation}

The evaluation emphasizes construct behavior, pattern analysis, robustness, and validation.

\paragraph{Controlled construct tests.}
We construct synthetic traces with total engagement fixed while varying when that engagement occurs, and test whether each measure responds to the temporal property it is designed to capture (RQ1--RQ3). 
Matched-volume traces compare activity with a concentrated burst. Permutations of an identical set of window-level values test the distinction between order-invariant continuity and order-sensitive recency: $P_h$ must remain constant under any rearrangement of the windows, whereas TH-GEE must change as high engagement moves toward or away from the most recent windows. 
Constructed dormant intervals test whether the reactivation measures separate continuous activity, first appearance, former activity, and renewed activity. Finally, a variant shifts publication times while holding engagement values fixed, checking that a late cluster of activity raises the recency-weighted score as intended.

\paragraph{Known-pattern topic analysis.}
Topics with an \emph{a priori} expected temporal signature (e.g., persistent, acute, cyclical, recent) provide a real-data construct test (RQ5): each measure should recover the pattern its topic is expected to display. The debunked-claim cohorts then test whether such topics share a common temporal profile or are heterogeneous (RQ6).

\paragraph{Reported quantities and robustness.}
For the topic analyses, we report active-window coverage, geometric continuity, signed recency lift, reactivation count, and reactivation-point engagement share. Because $P_h$ and TH-GEE are expressed on the engagement scale, cross-topic conclusions rely on the scale-free indicators (coverage, the recency ratio $\THGEE/P_h$, and $R_h$ with its reactivation count) together with topic comparisons and rank order, rather than raw-score magnitudes. Robustness (RQ4) is assessed by recalculating profiles across various window lengths and parameters. 
We report whether the qualitative interpretation and rank ordering of topics remain stable across these specifications, state the parameter values before presenting the corresponding results, and report alternative specifications rather than only those that produce the clearest separation.

\paragraph{Event-level validation.}
Daily Wikipedia pageview series for the same topics provide audience interactions dated to the day they occurred (RQ7). We recompute weekly traces from these series and compare the qualifying reactivation windows against the events located by two sequence-oriented methods on the same daily streams: Kleinberg's burst automaton \cite{kleinberg2003bursty} and PELT change-point estimation \cite{killick2012optimal}.

\subsection{Baselines and Parameter Selection}
\label{sec:parameters}

Aggregation baselines include the arithmetic mean and the median of the window-level scores. 
The unweighted geometric score is not an additional baseline: by Equation \eqref{eq:reduction}, it is identical to TH-GEE at $\gamma=0$. Sequence-oriented methods (e.g., Kleinberg burst scores, change-point statistics, Hawkes-process intensities \cite{kleinberg2003bursty,killick2012optimal,hawkes1971spectra}) require event-level interaction sequences that the publication data do not provide (Section~\ref{sec:sequence}); we therefore run them on the event-level Wikipedia series (Section \ref{sec:res_eventlevel}), where they serve as reference detectors for the reactivations identified by $R_h$.

Parameters are fixed by a documented procedure before results are inspected. Window lengths are drawn from a small set of substantively plausible values informed by prior attention-decay research \cite{wu2007novelty,crane2008robust}; the primary empirical specification uses 90-day (quarterly) windows. The activity threshold $q_h$ is set per topic to the 25th percentile of its nonzero window-level engagement scores, so that activity and dormancy are defined relative to each topic's own engagement scale rather than by a fixed absolute value. The primary empirical analysis uses decay coefficient $\gamma=0.1$ and dormancy duration $d=2$ (i.e., roughly half a year of inactivity on the quarterly grid); the synthetic construct tests use the parameter values stated alongside their results (e.g., $\gamma=0.5$, $q_h=5$, $d=3$ in Table~\ref{tab:construct}). Both $\gamma$ and $d$ are varied over a grid in the robustness checks. Applications that use TH-EE profiles as inputs to a downstream predictive task would additionally require selecting $\gamma$, $q_h$, and $d$ on training data and locking them before evaluating held-out traces; no such task is undertaken here.

\section{Results}
\label{sec:results}

The results are presented in three stages. First, controlled traces establish whether the measures exhibit their intended mathematical and construct properties. These analyses should be interpreted as construct checks rather than evidence of predictive performance.

Second, the topic-exemplar analysis examines whether the profiles of selected YouTube search cohorts are consistent with their anticipated persistent, acute, cyclical, or recent structure.

Third, the debunked-claim analysis examines heterogeneity and overlap in temporal-profile space. This analysis does not test whether TH-EE can classify misinformation. Rather, it tests whether attention associated with debunked claims can be reduced to a single characteristic temporal pattern.

\subsection{Construct validity of the Geometric Persistence Score (Addressing RQ1)}
\label{sec:res_persistence}

RQ1 asks whether $P_h$ separates engagement that is distributed across the horizon from matched-volume engagement concentrated in a single burst.
Table~\ref{tab:construct} reports the three measures for four canonical six-window patterns. The acute and continuous patterns are constructed to share the same total engagement (63 units, arithmetic mean $10.5$ per window), so sum- and mean-based summaries assign them identical scores, and a peak-based summary favors the burst. The Geometric Persistence Score instead separates them by an order of magnitude ($P_h = 1.000$ for the acute burst versus $P_h = 10.460$ for the continuous trace), because a single active window among five empty ones is heavily penalized by the geometric form. This result is consistent with the intended interpretation of geometric continuity: the score is reduced when engagement is concentrated in a small number of windows, even when total engagement is held constant.

\begin{table}[htbp]
\centering
\caption{Construct-validity values for four patterns (six windows).
Acute and continuous patterns share total volume (63) and arithmetic mean
($10.5$). $\operatorname{TH\mbox{-}GEE}$ uses $\gamma=0.5$; $R_h$ uses $q_h=5$,
$d=3$.}
\label{tab:construct}
\small
\renewcommand{\arraystretch}{1.2}
\begin{tabular}{lrrr}
\toprule
\textbf{Pattern (trace)} & $P_h$ & $\operatorname{TH\mbox{-}GEE}$ & $R_h$ \\
\midrule
Acute burst\\ \((63,0,0,0,0,0)\)        & 1.000  & 0.152  & 0.000 \\
Continuous\\ \((10,11,9,10,12,11)\)     & 10.460 & 10.846 & 0.000 \\
Formerly active\\ \((12,10,8,0,0,0)\)   & 2.298  & 0.529  & 0.000 \\
Re-emergent\\ \((12,10,0,0,0,14)\)      & 2.591  & 2.830  & 0.353 \\
\bottomrule
\end{tabular}
\end{table}

The remaining rows preview the complementary behavior of the other two measures.
The formerly-active and re-emergent traces have similar persistence ($P_h = 2.298$ and $2.591$) but sharply different recency ($\operatorname{TH\mbox{-}GEE} = 0.529$ versus $2.830$) and re-emergence ($R_h = 0.000$ versus $0.353$), illustrating that a single scalar cannot recover what the three-component profile encodes.

\subsection{Order sensitivity and recency weighting (Addressing RQ2)}
\label{sec:res_order}

RQ2 asks whether recency weighting introduces genuine order sensitivity that the unweighted score lacks. We enumerated all $720$ distinct permutations of the trace $(20,10,5,0,3,8)$. Across every permutation, the Geometric Persistence Score is constant to numerical precision (standard deviation below $10^{-15}$), confirming the permutation invariance of Equation~\ref{eq:permutation}. On the same permutations, $\operatorname{TH\mbox{-}GEE}$ with $\gamma=0.5$ varies substantially (standard deviation $2.04$; range $[1.89, 10.48]$): the score is lowest when large values sit in early windows and highest when they sit in recent windows. We also verified the reduction identity of Equation \ref{eq:reduction} directly: across permutations, the maximum absolute difference between $P_h$ and $\operatorname{TH\mbox{-}GEE}(\gamma=0)$ is below $10^{-14}$. The unweighted geometric score and TH-GEE at $\gamma=0$ are therefore the same estimator, and they must not be presented as separate baselines.

The direction of the recency effect is also as intended. For an early-loaded trace $(20,12,8,5,2,1)$, increasing $\gamma$ from $0$ to $2$ drives $\operatorname{TH\mbox{-}GEE}$ monotonically down from $5.68$ to $1.14$; for the mirror-image late-loaded trace $(1,2,5,8,12,20)$, the same sweep drives it monotonically up from $5.68$ to $18.52$. Both coincide at $\gamma=0$, where the measure reduces to $P_h$.

\subsection{Re-emergence construct test (Addressing RQ3)}
\label{sec:res_reemergence}

RQ3 asks whether $R_h$ separates continuously active engagement from engagement that returns after dormancy. In Table~\ref{tab:construct}, the continuous trace receives $R_h = 0$ because it never becomes dormant, and the formerly-active trace receives $R_h = 0$ because activity does not return; only the re-emergent trace, which becomes active again after three dormant windows, receives a positive score ($R_h = 0.353$), with the reactivation correctly localized to the final window. The ``prior activity'' condition in Equation~\ref{eq:reactivation} is likewise necessary: a late first appearance such as $(0,0,0,0,0,14)$ yields $R_h = 0$, so a campaign's initial emergence is never mislabeled as re-emergence.

\subsection{Parameter sensitivity and temporal baselines (Addressing RQ4)}
\label{sec:res_sensitivity}

RQ4 concerns sensitivity to the decay coefficient $\gamma$, the dormancy duration $d$, and the activity threshold $q_h$. The $\gamma$ sweeps in Section~\ref{sec:res_order} already show the expected monotone behavior with a clear direction determined by where engagement mass sits. Re-emergence responds to $d$ as a graduated dormancy filter: for the trace $(12,10,0,14,0,0,0,15)$ with $q_h=5$, requiring $d=1$ dormant window detects both returns ($R_h = 0.525$), $d=2$ or $d=3$ retains only the return that follows the longer gap ($R_h = 0.265$), and $d=4$ suppresses both ($R_h = 0$). The activity threshold $q_h$ behaves as a reclassification control: raising it changes which windows count as active and can both create and remove qualifying dormant periods, so $R_h$ is not monotone in $q_h$ and, once $q_h$ exceeds the peak of a trace, drops to zero. These results reinforce the paper's methodological recommendation to report full sensitivity surfaces and to lock parameters on held-out folds rather than reporting a single favored setting. 

\subsection{Known-pattern topic analysis (Addressing RQ5)}
\label{sec:res_observational}

\begin{figure*}[htbp]
\centering
\includegraphics[width=0.8\textwidth]{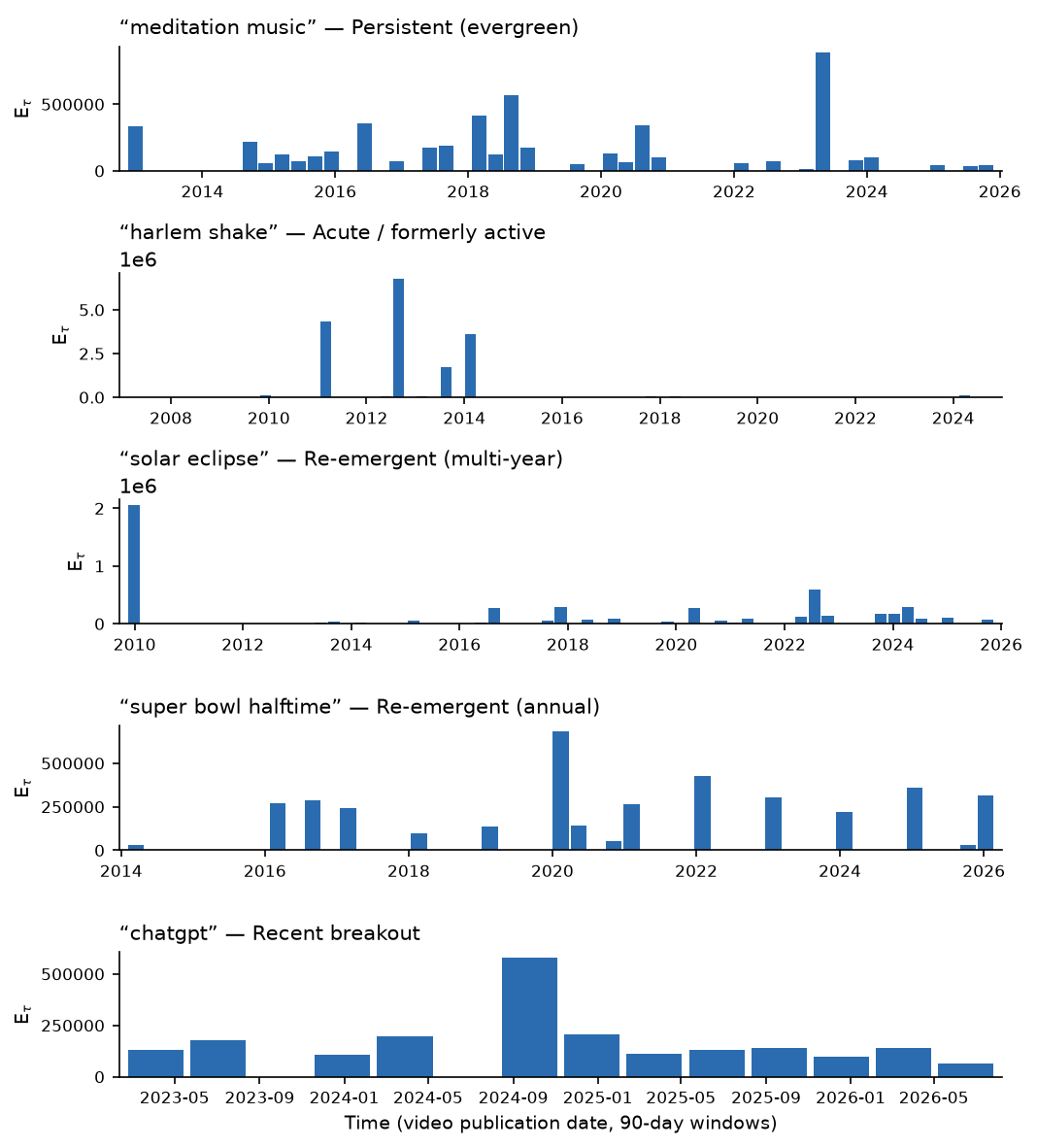}
\caption{Quarterly weighted-engagement traces $E_\tau$ for five featured topics. The evergreen topic is spread across the whole horizon; the fad is a short early burst followed by silence; the two cyclical topics recur at multi-year and annual periods; the recent breakout is confined to the most recent windows}
\label{fig:exemplars}
\end{figure*}

We now ask whether the measures recover, on real data, the temporal signature each topic is expected to display. Table~\ref{tab:youtube} reports the profile for five featured topics, and Figure~\ref{fig:exemplars} shows their quarterly traces. Because $P_h$ and $\operatorname{TH\mbox{-}GEE}$ are on the engagement scale, and viral topics can reach very large magnitudes, cross-topic reading relies on the scale-free columns: active-window coverage, the recency ratio $\operatorname{TH\mbox{-}GEE}/P_h$, and $R_h$ with its reactivation count.

\begin{table*}[!htbp]
\centering
\caption{Temporal profiles for five YouTube topics with known expected patterns (90-day windows, up to 50 videos each). Coverage is the fraction of windows that are active; ``recency'' is $\operatorname{TH\mbox{-}GEE}/P_h$ (${>}1$ means recency-weighted); ``react.'' is the number of qualifying reactivations. $P_h$ and $\operatorname{TH\mbox{-}GEE}$ (recency) magnitudes are on the engagement scale and are not comparable across topics; the scale-free columns are.}
\label{tab:youtube}
\small
\renewcommand{\arraystretch}{1.2}
\begin{tabular}{llrrrrrr}
\toprule
\textbf{Topic} & \textbf{Expected} & \textbf{span} & $P_h$ & \textbf{coverage} & \textbf{recency} & $R_h$ & \textbf{react.} \\
\midrule
meditation music     & persistent          & 13.0\,yr & 594.6   & 55\% & 0.78 & 0.145 & 4 \\
harlem shake         & acute / former      & 17.6\,yr & 11.6    & 22\% & 1.26 & 0.315 & 5 \\
solar eclipse        & re-emergent (multi) & 15.8\,yr & 89.3    & 40\% & 8.60 & 0.279 & 7 \\
super bowl halftime  & re-emergent (annual)& 12.0\,yr & 50.9    & 33\% & 2.38 & 0.523 & 8 \\
chatgpt              & recent              & 3.4\,yr  & 26976.3 & 86\% & 1.56 & 0.085 & 1 \\
\bottomrule
\end{tabular}
\end{table*}

The profiles recover the expected patterns. The evergreen topic \emph{meditation music} is active across most of a 13-year horizon ($55\%$ coverage), with a balanced recency ratio ($0.78$) and little re-emergence, the signature of sustained, order-insensitive activity. The dead fad \emph{harlem shake} is the opposite: its activity is confined to a short early burst (Figure~\ref{fig:exemplars}), giving low coverage ($22\%$) and, on a quarterly grid, essentially no engagement in the final windows. The two cyclical topics are separated from both by their re-emergence: \emph{super bowl halftime} returns almost every year ($R_h = 0.523$, eight reactivations) and \emph{solar eclipse} returns at the 2017 and 2024 eclipses (seven reactivations) with a strongly recency-weighted profile (ratio $8.6$) driven by the large 2024 spike. The recent breakout \emph{chatgpt} is active throughout its short post-2022 horizon with a recency ratio above one and negligible re-emergence ($R_h = 0.085$): it has been continuously present since launch and has not yet gone dormant. No single scalar reproduces these distinctions; the three-component profile does.

Across the full set of eighteen topics, the same separation holds in aggregate: the persistent topics have the highest coverage and near-balanced recency, the acute topics the lowest coverage, and the cyclical topics the largest reactivation counts. Consistent with the framework's caution, $P_h$ magnitude tracks engagement volume (the recent viral topic \emph{chatgpt} attains $P_h \approx 2.7\times10^{4}$ purely from its enormous per-video engagement), which is exactly why cross-topic claims should rest on the scale-free indicators rather than on $P_h$ itself.

\subsection{Application to debunked-claim search cohorts (Addressing RQ6)}
\label{sec:res_debunked}

We next examine 19 search cohorts associated with claims that have been authoritatively debunked. These queries return a mixture of content (e.g., endorsement, criticism, fact-checking, news coverage, commentary, parody, entertainment). The resulting traces, therefore, measure observable attention to a topic, but not belief in or acceptance of the associated claim.

The purpose of the analysis is not to compare ``misinformation'' with ``benign'' topics. Instead, it asks whether public attention organized around debunked claims exhibits a common temporal structure. If such cohorts occupy a narrow and distinctive region of profile space, temporal structure might provide a useful group-level characterization. If they are widely distributed and overlap with comparison topics, temporal heterogeneity is the more appropriate substantive finding.

Under the primary parameter specification, all 19 retrieved cohorts contain at least two qualifying reactivation points. This result should not be interpreted as evidence that all misinformation universally re-emerges. It applies to the sampled search cohorts within the stated window, activity threshold, and dormancy definitions. The magnitude and frequency of reactivation nevertheless vary substantially across cohorts.

\begin{table*}[!htbp]
\centering
\caption{Temporal profiles for nineteen debunked-claim topics (90-day windows).
Columns are as in Table~\ref{tab:youtube}. The profile column is a
\emph{descriptive} summary of the two scale-free columns that vary across this
set, assigned after measurement and stated in the framework's own terms:
\emph{persistent} versus \emph{intermittent} is active-window coverage at or
below $55\%$, and \emph{recency-weighted} versus \emph{strongly
recency-weighted} is a recency ratio at or below $10$. The bands are reading
aids (not classifiers) and no topic is assigned to a re-emergence category
because all nineteen are re-emergent ($R_h$ from $0.093$ to $0.437$, at least
two qualifying reactivations each); the $R_h$ and react.\ columns carry that
magnitude directly.}
\label{tab:debunked}
\small
\renewcommand{\arraystretch}{1.2}
\begin{tabular}{>{\raggedright\arraybackslash}p{0.19\linewidth}>{\raggedright\arraybackslash}p{0.23\linewidth}rrrrr}
\toprule
\textbf{Topic} & \textbf{Profile} & \textbf{span} & \textbf{coverage} & \textbf{recency} & $R_h$ & \textbf{react.} \\
\midrule
flat earth                         & persistent, strongly recency-weighted   & 11.6\,yr & 67\% & 15.3 & 0.100 & 3 \\
pizzagate                          & intermittent, recency-weighted      & 9.5\,yr & 44\% & 3.1 & 0.323 & 5 \\
5G--coronavirus                    & intermittent, recency-weighted      & 6.2\,yr & 46\% & 1.4 & 0.277 & 3 \\
moon-landing hoax                  & intermittent, strongly recency-weighted & 17.7\,yr & 36\% & 44.5 & 0.247 & 6 \\
chemtrails                         & intermittent, strongly recency-weighted & 15.7\,yr & 42\% & 12.6 & 0.236 & 6 \\
bigfoot                            & intermittent, strongly recency-weighted & 18.1\,yr & 32\% & 67.4 & 0.224 & 5 \\
Loch Ness monster                  & intermittent, strongly recency-weighted & 14.9\,yr & 43\% & 23.6 & 0.159 & 4 \\
Area 51 aliens                     & intermittent, recency-weighted      & 11.2\,yr & 52\% & 6.6 & 0.262 & 6 \\
climate change (hoax)              & intermittent, strongly recency-weighted & 16.6\,yr & 40\% & 21.1 & 0.313 & 8 \\
Mandela effect                     & persistent, recency-weighted        & 9.8\,yr & 60\% & 3.3 & 0.129 & 3 \\
Birds aren't real                  & persistent, recency-weighted        & 8.6\,yr & 74\% & 2.5 & 0.093 & 2 \\
Hollow Earth                       & intermittent, strongly recency-weighted & 17.5\,yr & 40\% & 36.9 & 0.108 & 3 \\
phantom time hypothesis            & intermittent, strongly recency-weighted & 13.1\,yr & 44\% & 15.6 & 0.256 & 6 \\
Project Blue Beam                  & intermittent, strongly recency-weighted & 17.4\,yr & 25\% & 27.7 & 0.228 & 4 \\
HAARP weather control              & intermittent, strongly recency-weighted & 19.1\,yr & 46\% & 11.1 & 0.284 & 9 \\
Paul McCartney (``Paul is dead'')  & intermittent, recency-weighted      & 19.0\,yr & 41\% & 8.6 & 0.209 & 6 \\
fluoride water poisoning           & intermittent, strongly recency-weighted & 19.7\,yr & 39\% & 14.1 & 0.437 & 13 \\
Mothman                            & persistent, recency-weighted        & 11.4\,yr & 62\% & 2.9 & 0.289 & 8 \\
CERN portals                       & intermittent, strongly recency-weighted & 16.2\,yr & 32\% & 66.5 & 0.196 & 4 \\
\bottomrule
\end{tabular}
\end{table*}

\begin{figure*}
    \centering
    \includegraphics[width=0.9\linewidth]{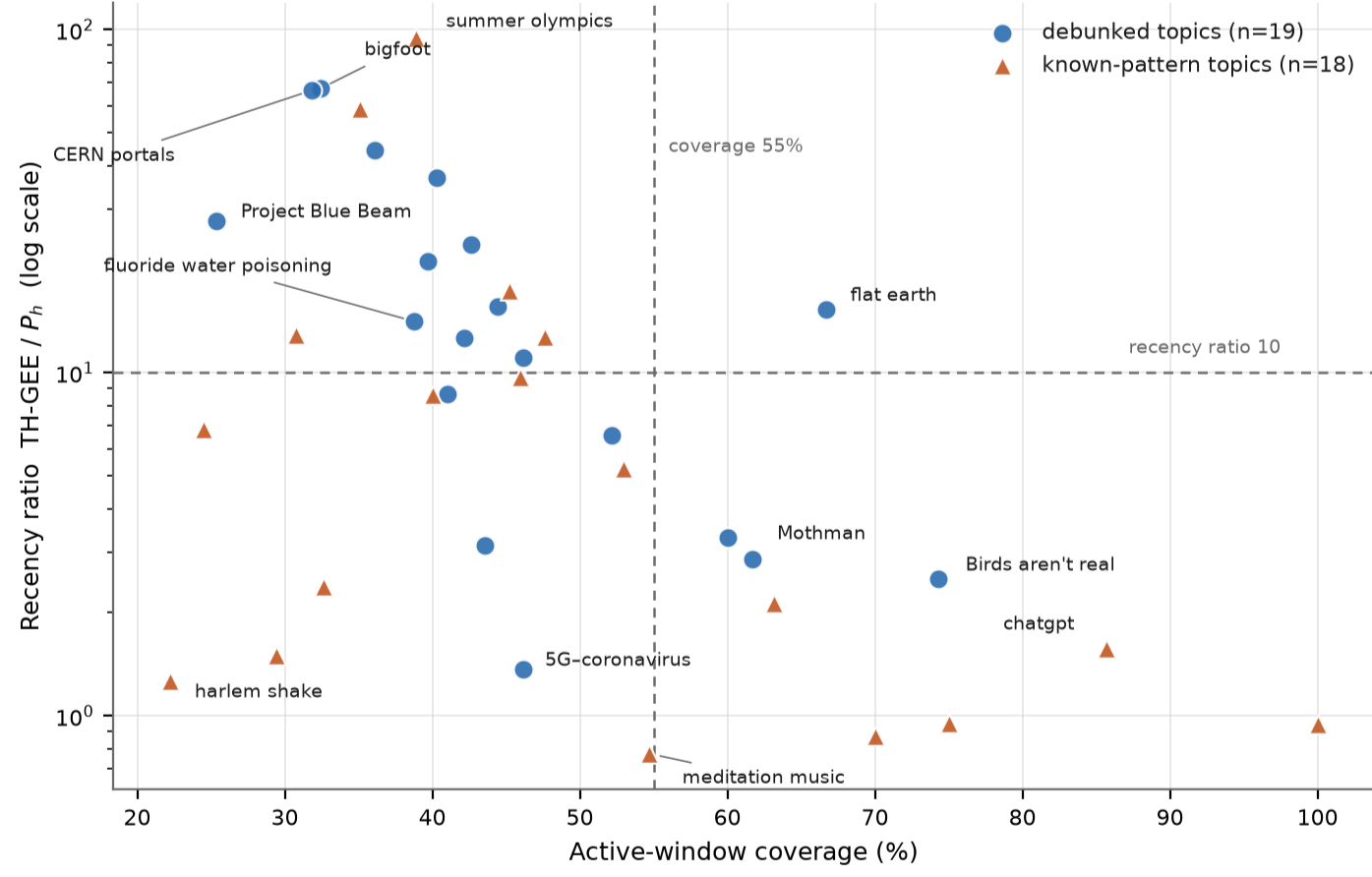}
    \caption{Distribution of 18 comparison-topic cohorts and 19 debunked-claim search cohorts in temporal-profile space. The horizontal axis reports active-window coverage, and the vertical axis reports signed recency lift. The groups overlap substantially, indicating that the retrieved debunked-claim cohorts do not form a distinct temporal cluster. The figure describes the sampled search cohorts and is not a classifier of content veracity or user belief}
    \label{fig:debunkedspace}
\end{figure*}

The central finding is not merely that temporal measures fail to separate debunked claims from benign topics. More importantly, the debunked-claim topics themselves \textit{do not share a common temporal structure}. They occupy multiple regions of the temporal-profile space, spanning coverage and recency ranges comparable to those of the known-pattern topics, depicted in  Figure \ref{fig:debunkedspace}.

The dataset separates along two of the three profile axes. On the persistence axis, coverage ranges from $25\%$ (\emph{Project Blue Beam}) to $74\%$ (\emph{Birds aren't real}): the most distributed topics are active across roughly three-quarters of a horizon spanning years, while the most concentrated are active in a quarter of theirs. 
On the recency axis, the ratio $\operatorname{TH\mbox{-}GEE}/P_h$ ranges from $1.4$ to $67.4$, and the two axes are clearly independent. \emph{Bigfoot} spans $18.1$ years, among the longest horizons in the table, yet carries a ratio of $67.4$: an old claim whose engagement mass sits almost entirely in recent windows. \emph{5G--coronavirus}, by contrast, has the shortest span in the set ($6.2$ years, consistent with a claim that originated in a discrete 2020 event) but a nearly balanced ratio of $1.4$. A recently \emph{originating} topic is therefore not the same as a recency-\emph{weighted} one, and the profile distinguishes the two.

The third axis does not separate this set: every one of the nineteen topics is re-emergent, with at least two qualifying reactivations and $R_h$ between $0.093$ and $0.437$. Re-emergence here is a shared property rather than a discriminating one, and its magnitude still varies more than fourfold across the set: \emph{fluoride water poisoning} records the strongest re-emergence ($R_h = 0.437$, thirteen reactivations), against \emph{Birds aren't real} at $R_h = 0.093$ with two. 

Notably, every topic we examined retains recent engagement. The one known pattern absent from the set is the acute, formerly active signature of a dead fad: no debunked topic here shows the low coverage combined with negligible terminal activity that characterizes \emph{harlem shake} in Table~\ref{tab:youtube}. 

Overall, these results indicate that misinformation-related attention is temporally heterogeneous. 
Within the topics observed here, a debunked claim may remain broadly distributed across its horizon or be confined to a fraction of it. The topic may also carry a nearly balanced recency profile or one dominated by the most recent windows, and may return after dormancy either occasionally or repeatedly.

The distinction of patterns has practical importance. A persistent false claim may require sustained educational or pre-bunking efforts; an acute claim may require immediate surge response; a dormant claim that re-emerges may require renewed contextualization; and an old claim with a sharply increasing recency profile may signal that a contemporary event, influencer, or platform mechanism has restored its relevance. The framework therefore contributes not by assigning a temporal signature to misinformation as a class, but by identifying the different temporal states through which misinformation narratives can move.

\subsection{Event-level validation on Wikipedia pageviews (Addressing RQ7)}
\label{sec:res_eventlevel}

\begin{figure*}
    \centering
    \includegraphics[width=0.75\linewidth]{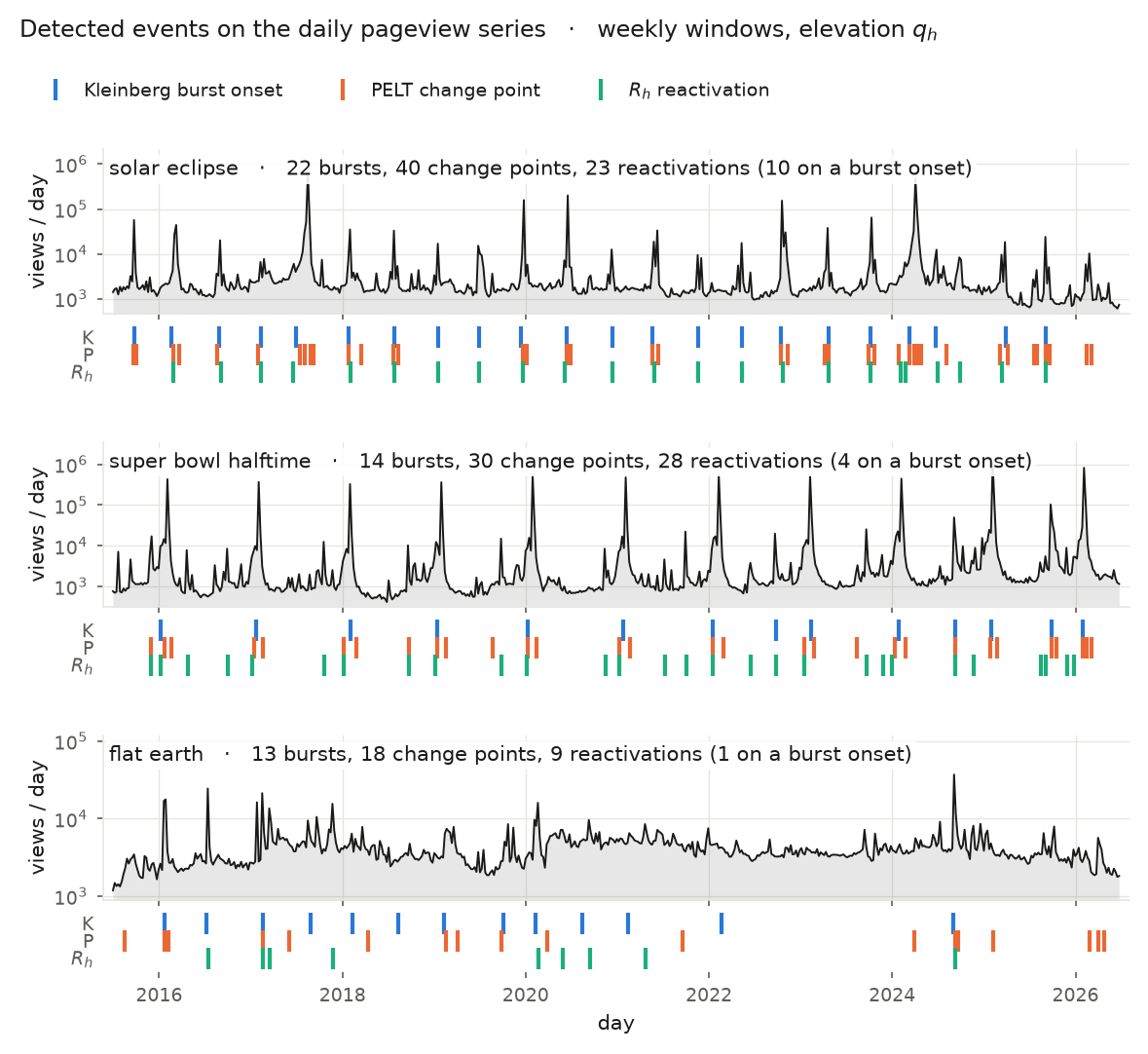}
    \caption{Event-level validation on daily Wikipedia pageview series for three of the study topics, 2015-07-01 to 2026-06-30. Each panel plots the weekly maximum of the daily series on a logarithmic scale. The strip beneath each panel marks the events located by Kleinberg's burst automaton (K) \cite{kleinberg2003bursty}, by PELT change-point estimation (P) \cite{killick2012optimal}, and the qualifying reactivation windows identified by $R_h$ on a weekly grid. Because a continuously read article never falls below a quantile of its own distribution, $q_h$ here is twice each topic's median window score. Across all 37 topics, reactivation windows contain a burst onset $10.7\times$ and a change point $6.6\times$ more often than a uniformly chosen window (permutation $p=0.0001$). The measures agree on the recurring structure: the annual \emph{super bowl halftime} peaks and the eclipse spikes in \emph{solar eclipse}. $R_h$ registers only a subset of what the event-level detectors find.}
    \label{fig:eventlevel}
\end{figure*}

Because the YouTube API provides publication timestamps with collected engagement totals rather than repeated engagement snapshots, we treat the re-emergence results above as a construct demonstration over topic activity. Event-level validation is nevertheless possible on a second data source. The Wikimedia REST API serves daily pageview counts per article from 2015 onward, which are audience interactions dated to the day they occurred rather than publication cohorts, and 37 of our topics have a corresponding English Wikipedia article. Recomputing weekly traces from those series over 2015-07-01 to 2026-06-30 and running two sequence-oriented baselines on the same daily streams (Kleinberg's burst automaton \cite{kleinberg2003bursty} and PELT change-point estimation \cite{killick2012optimal}) yields the comparison shown in Figure \ref{fig:eventlevel}.

Qualifying reactivation windows contain a burst onset $10.7$ times more often, and a change point $6.6$ times more often, than a uniformly chosen window (permutation $p=0.0001$ in both cases), and across the 37 topics the reactivation count correlates with the change-point count at Spearman $\rho = 0.56$. One adjustment is required: because $q_h$ is a quantile of a topic's own distribution, a continuously read Wikipedia article never falls below it and never registers as dormant, so for this data $q_h$ is set to twice each topic's median window score, making ``active'' mean elevated above baseline rather than above the lower quartile. Reactivation windows capture 20\% of burst onsets and 13\% of change points, which supports the positioning in Section \ref{sec:attention}: $R_h$ is a coarser and more selective instrument that complements these methods rather than replacing them. Hawkes-process intensities \cite{hawkes1971spectra} require individual event marks rather than daily aggregates and remain future work.

\section{Discussion}
\label{sec:discussion}

\subsection{Findings}
\label{sec:findings}

This paper draws five findings.
First, the empirical profiles dissociate persistence, recency, and re-emergence, confirming that the three dimensions measure distinct properties of the same traces. Persistence does not imply recency: \emph{meditation music} is active across 55\% of a 13-year horizon with a balanced recency ratio (0.78), whereas \emph{bigfoot} is active in only 32\% of its windows yet carries a ratio of 67.4, with its engagement mass concentrated almost entirely in recent windows. Recency does not imply re-emergence: \emph{chatgpt} is recency-weighted (ratio 1.56) but records almost no re-emergence ($R_h=0.085$) because it has been continuously active since first appearing, whereas \emph{super bowl halftime} returns from dormancy almost annually ($R_h=0.523$, eight reactivations). And persistence does not imply re-emergence: \emph{Birds aren't real}, the most broadly distributed debunked topic (74\% coverage), is also the least re-emergent ($R_h=0.093$), while \emph{fluoride water poisoning} combines low coverage (39\%) with the strongest re-emergence in the set ($R_h=0.437$, thirteen reactivations).

Second, temporal distribution and engagement magnitude should be reported separately. Active-window coverage provides a scale-free description of how widely activity is distributed, whereas geometric continuity retains information about engagement magnitude. Signed recency lift then describes how the temporal placement of engagement differs from its unweighted distribution. This separation improves interpretability and reduces the risk of labeling a high-volume topic as more persistent solely because its engagement values are larger.

Third, the debunked-claim search cohorts are temporally heterogeneous. They overlap with comparison topics and vary in coverage, recency, and reactivation. The substantive implication is not that temporal measures are ineffective. Rather, it is that misinformation-related attention should not be assumed to possess a universal temporal signature.

Fourth, the event-level validation shows both convergence and a boundary. On daily pageview series, the reactivations identified by $R_h$ co-locate with burst onsets and change points far more often than chance, yet capture only a fraction of the events those detectors find, and topic-level agreement between publication-cohort and event-level profiles is weak for persistence and re-emergence. Publication-cohort traces are therefore an informative proxy for how distributed and how recent attention is, but not for whether it returned; re-emergence claims should rest on event-level data where available.

Fifth, temporal profiles may still be analytically useful once a topic has been identified through independent content, contextual, network, or provenance evidence. Persistent attention may motivate longitudinal study; a sharp positive recency lift may identify a topic gaining contemporary salience; and reactivation following dormancy may direct attention toward a triggering event or renewed production episode. These interpretations concern the state of observable attention, not its cause or social effect.

\subsection{Why misinformation exhibits heterogeneous temporal signatures}
\label{sec:misinformation_mechanisms}

The third finding of Section~\ref{sec:discussion} raises an immediate question: \textit{if debunked-claim attention has no universal temporal signature, what produces the variety of signatures it does have?} The observed traces alone cannot identify the causal mechanism responsible for each trajectory. The YouTube data record when videos were published and their eventual engagement totals; they do not reveal \textit{why} a user produced, encountered, or engaged with a particular item. This section therefore identifies theoretically plausible mechanisms consistent with the observed patterns, without claiming that TH-EE establishes any of them.

\paragraph{Triggering events.}
Acute and re-emergent patterns may be produced by changes in the surrounding information environment, caused by certain events. 
A crisis, scientific announcement, political controversy, anniversary, media release, or statement by a prominent actor can abruptly increase the relevance of an existing narrative. Research on collective attention distinguishes attention generated by endogenous diffusion from activity driven by exogenous events and shows that external stimuli frequently account for sudden peaks in online discussion \cite{crane2008robust,lehmann2012dynamical}. 
For example, the 5G--coronavirus topic is consistent with a discrete external trigger because the claim became salient alongside the emergence of COVID-19 in 2020. 
A moon-landing hoax topic, by contrast, may regain salience around certain anniversaries, space missions, documentaries, films, or public discussions. Recurrence does not require continuous engagement or a coordinated campaign; a new event can make a dormant claim relevant again.

\paragraph{Narrative durability and adaptability.}
Some misinformation topics remain available for repeated use because they concern durable objects of suspicion, unresolved mysteries, institutional distrust, or identity-relevant beliefs. 
These narratives can be adapted to new evidence, actors, technologies, or political conditions without being replaced by a new claim. A persistent topic, such as flat earth, is supported by an enduring community and a large body of reusable demonstrations, rebuttals, humorous content, and counterarguments. 
Other narratives may remain dormant until they can be connected to a new event. Research on misinformation lifecycles has similarly shown that rumors can mutate over time and that their temporal behavior varies with message characteristics and sources \cite{shin2018diffusion}. Recent evidence suggests that false and ambiguous stories are more likely to reappear online, but this is a statistical tendency rather than a universal pattern \cite{kauk2025reappearance}.

\paragraph{Different motivations for engagement.}
Topic-level engagement is generated by more than acceptance of the underlying misinformation. Users may engage because they believe the claim, but they may also criticize it, debunk it, ridicule it, investigate it, or use it as entertainment. Outrage, disagreement, identity signaling, and out-group conflict can themselves produce substantial interaction \cite{brady2017emotion}. 
Misinformation has also been found to benefit disproportionately from outrage-related reactions \cite{rathje2024outrage}. Consequently, continuing attention does not necessarily indicate continuing persuasion. A controversy may persist because believers, skeptics, journalists, educators, and entertainers repeatedly interact around the same topic for different reasons.

These mixed motivations help explain why debunked claims can resemble benign content. Both conspiracy theories and sporting events can generate anticipation, argument, humor, community participation, commentary, and reaction content. The observable temporal signature records the resulting attention but does not distinguish the psychological or communicative \textit{motive} behind each interaction.

\paragraph{Communities, habits, and repeated production.}
Persistence may also arise when a stable community repeatedly produces and interacts with topic content. Engagement can become habitual rather than being generated by a new evaluation of each claim \cite{ceylan2023habitual}.
Within ideologically or topically aligned communities, repeated exposure and social reinforcement can sustain activity even when the topic receives little attention from the wider public \cite{cinelli2021echo}. A persistent trace may represent many small acts of routine production and engagement rather than a continuous viral cascade.
The size, commitment, and production practices of the relevant community affect the resulting pattern. A large, durable community can support continuous activity. A smaller community may generate intermittent clusters. A topic dependent on a few prominent creators may vary with their publication schedules.

\paragraph{Platform visibility and algorithms.}
Observed temporal patterns are also shaped by how platforms rank, recommend, search, and resurface content. Because user attention is limited, the probability of engagement depends partly on whether content remains visible when a user is present \cite{hodas2014simple}. Platform algorithms also treat false or misleading content differently, while user behaviors such as superspreading further shape diffusion across the network. Recommendation systems can connect users to older material, reinforce repeated exposure, and influence how misinformation producers reach and interact with audiences \cite{pathak2023recommendation}. These systems may also amplify sensational or divisive content because such material often attracts more clicks and reactions \cite{pathak2023recommendation}. Importantly, these mechanisms operate on both benign and misleading content. The same recommendation process may resurface an old conspiracy-theory video, an eclipse documentary, a music performance, or an educational tutorial in response to similar engagement and relevance signals.

\paragraph{Strategic amplification and counter-messaging.}
In some cases, temporal behavior may reflect deliberate intervention. Coordinated actors can sustain a narrative through repeated posting, introduce modified variants, or reactivate an older claim when conditions become favorable \cite{disarm2024framework,rushing2026characterising,rushing2026cogwarfare}. Such activity interacts with ordinary diffusion processes, including repeated sharing, influential accounts, and community-based propagation \cite{bakshy2011everyone,delvicario2016spreading,shin2018diffusion,vosoughi2018spread}. False and ambiguous stories may also be more likely to reappear \cite{kauk2025reappearance}.

Conversely, journalists, fact-checkers, institutions, and other users may generate renewed topic attention through correction and counter-messaging. Corrective interventions can reduce subsequent engagement with and diffusion of false information \cite{slaughter2025community}. However, the publication of corrections may itself contribute additional observable content and discussion about the topic. The resulting trace may therefore contain both attempted amplification and attempted mitigation. This mechanism is especially important when interpreting re-emergence. TH-EE can identify that renewed activity occurred, but distinguishing deliberate amplification, organic discussion, correction, and other explanations requires content analysis, account- and network-level evidence, provenance information, and event-level investigation \cite{wardle2017information,disarm2024framework,rushing2026characterising}.

\smallskip
Taken together, these mechanisms help explain why misinformation does not possess a universal temporal signature. The same forces that shape misleading content (e.g., external events, community interest, repeated production, and platform visibility) also shape benign attention \cite{shin2018diffusion,hodas2014simple,pathak2023recommendation}.

\subsection{Interpreting and Applying Temporal Profiles}
\label{sec:practical_implications}

The practical value of the temporal profile lies not only in the distinctions it identifies, but also in the inferences it prevents. A high persistence score does not imply that content is coordinated, adversarial, synthetic, or persuasive. Benign evergreen material, including health guidance, institutional communication, seasonal topics, and durable online communities, can produce the same temporal signature as strategically sustained messaging.

The YouTube topics illustrate this directly: the most persistent and the most re-emergent topics here are ordinary cultural and calendar phenomena (e.g., meditation music, eclipses, the Super Bowl) not influence operations. The debunked-claim topics of Section~\ref{sec:res_debunked} make the point sharper still: they span the same range of temporal patterns as the benign topics, so even in the domain where a temporal shortcut would be most welcome, no signature marks a claim as false. The measures best serve as triage and description instruments. They can prioritize traces for analyst attention and compare the temporal structure of narratives, but the inferential step from temporal structure to intent requires further evidence.

Two measurement points deserve emphasis. First, because $P_h$ and $\operatorname{TH\mbox{-}GEE}$ are expressed on the engagement scale, their magnitudes are not comparable across corpora or eras with different engagement levels; comparisons across such boundaries should rely on scale-free quantities such as the re-emergence fraction, rank structure, and permutation-based order sensitivity, as we do here. 
Second, the equivalence of the unweighted geometric score and $\operatorname{TH\mbox{-}GEE}$ at $\gamma=0$ is not a limitation to be worked around but a consistency pattern to be preserved: any implementation or results table in which the two differ contains an error. The reference implementation and analysis code accompanying this article are structured so that these identities are exercised by tests, which we recommend as a reproducibility check for future extensions of the framework.

\subsection{Study Validity}
\label{sec:limitations}

\paragraph{Construct validity.} TH-EE measures observable publication and engagement patterns, not belief, persuasion, endorsement, truth, harmfulness, or intent. Likes and comments may reflect approval, criticism, correction, curiosity, identity expression, or entertainment. Temporal activity should therefore not be interpreted as a direct measure of cognitive or social effect.

\paragraph{Measurement validity.} The YouTube application uses publication timestamps and cumulative interaction counts observed at collection time. It does not contain event-level timestamps for individual likes and comments. The resulting traces characterize publication cohorts and their eventual observed engagement rather than the actual timing of audience interactions. Older videos have also had more time to accumulate interactions than recently published videos.

An event-level check on daily Wikipedia pageview series for the same topics indicates that this substitution matters unevenly: the group-level result of Section \ref{sec:res_debunked} replicates, but topic-level agreement between publication-cohort and event-level profiles is strong for neither persistence nor re-emergence, so publication-cohort traces should be read as a proxy for how distributed and how recent attention is rather than for whether it returned.

\paragraph{Sampling validity.} Each topic cohort contains up to 50 videos returned for a search query. Search results may be influenced by query wording, indexing, relevance ranking, platform changes, availability, and collection date. The sample is neither a census nor a probability sample, and sparse retrieval across a long time horizon may affect measured coverage, dormancy, and reactivation.

\paragraph{Parameter dependence.} The resulting profile depends on the observation horizon, window length, recency parameter, activity threshold, and dormancy requirement. These choices are substantive modeling decisions rather than universally correct constants. Robustness analyses can reveal whether conclusions depend on a narrow specification, but they cannot establish one parameter set as optimal for all platforms or topics.

\paragraph{External validity.} The empirical application is limited to YouTube search cohorts and Wikipedia pageview events. Platforms differ in content format, recommendation systems, interaction mechanisms, audience composition, and data availability. Cross-platform validation on social media interaction data is required before generalizing the profile distributions reported here.

\section{Conclusion}
\label{sec:conclusion}

This paper introduces a framework for describing three temporal dimensions of online attention: persistence, recency, and re-emergence. Rather than reducing temporal behavior to a single engagement total or class label, TH-EE reports a profile containing activity coverage, geometric continuity, signed recency lift, reactivation frequency, and reactivation-point engagement share.

Controlled analyses demonstrate the distinct behavior of these measures. Coverage and geometric continuity distinguish broadly distributed activity from concentrated bursts; recency weighting introduces sensitivity to temporal order; and the reactivation measures distinguish renewed activity from continuous activity and first appearance.

The YouTube application provides a proof of concept using publication-cohort engagement traces. Topics selected as persistent, acute, cyclical, and recent exemplars display corresponding temporal characteristics. Search cohorts associated with debunked claims, however, do not exhibit a single temporal signature. They vary across the profile dimensions and overlap substantially with comparison topics. An event-level validation on daily Wikipedia pageview series confirms that the framework's reactivations align with independently detected bursts and change points, while showing that publication-cohort traces proxy distribution and recency more reliably than re-emergence.

This heterogeneity is a substantive result. It indicates that attention to misinformation-related topics cannot be identified or explained from timing alone. Temporal measurement instead supports a narrower question: whether an identified topic is broadly persistent, increasingly salient, dormant, or returning after dormancy. Answering why that state occurred requires additional content, contextual, network, and behavioral evidence.

Future research should extend the event-level validation to individual-event streams suitable for Hawkes-process modeling, to repeated platform observations, to probability-based or substantially larger samples, and to cross-platform social media data. Such studies could investigate how temporal profiles relate to external events, community structure, platform visibility, corrective interventions, and changes in public attention.

\backmatter

\bmhead{Acknowledgments} The views expressed are those of the authors and do not reflect the official policy or position of the United States Air Force, the Department of Defense, or the United States Government.

\section*{Statements and Declarations}

\bmhead{Funding}
The authors did not receive support from any organization for the submitted work.

\bmhead{Competing Interests}
On behalf of all authors, the corresponding author states that there is no conflict of interest.

\bmhead{Data Availability}
All materials needed to reproduce the empirical results are provided as a self-contained reproducibility bundle deposited at \url{https://github.com/bnatc85/TH-GEE}. The bundle contains the reference implementation of the three measures, the analysis and figure scripts, and a frozen snapshot of the collected YouTube metadata. Raw metadata were collected via the YouTube Data API v3 under its standard terms of service; because live search results change over time, the frozen snapshot is the authoritative artifact.
The bundle also contains the event-level validation of Section \ref{sec:res_eventlevel}: the topic-to-article mapping, a frozen snapshot of the daily Wikipedia pageview series, and reference implementations of the two sequence-oriented baselines with their construct tests.

\bibliography{references_JCSS}
\end{document}